\documentclass[sigconf]{acmart}
\usepackage[ruled]{algorithm2e}
\usepackage{multirow}
\usepackage{subfigure}
\usepackage{enumitem}
\usepackage{sistyle}
\SIthousandsep{,}
\usepackage{makecell}
\usepackage{multirow}
\usepackage{enumitem}
\usepackage{filecontents}
\usepackage{bbm}
\usepackage{pgfplots}
\usetikzlibrary{patterns}

\definecolor{c1}{RGB}{255,255,157}
\definecolor{c2}{RGB}{190, 235, 159}
\definecolor{c3}{RGB}{94,180,210}

\definecolor{c4}{RGB}{18, 50, 8}

\AtBeginDocument{%
  \providecommand\BibTeX{{%
    \normalfont B\kern-0.5em{\scshape i\kern-0.25em b}\kern-0.8em\TeX}}}

\copyrightyear{2021} 
\acmYear{2021} 
\setcopyright{acmcopyright}\acmConference[CIKM '21]{Proceedings of the 30th ACM International Conference on Information and Knowledge Management}{November 1--5, 2021}{Virtual Event, QLD, Australia}
\acmBooktitle{Proceedings of the 30th ACM International Conference on Information and Knowledge Management (CIKM '21), November 1--5, 2021, Virtual Event, QLD, Australia}
\acmPrice{15.00}
\acmDOI{10.1145/3459637.3482345}
\acmISBN{978-1-4503-8446-9/21/11}

\makeatletter
\g@addto@macro\normalsize{%
  \abovedisplayskip 1pt plus1pt 
  \belowdisplayskip 1pt plus1pt
  \abovedisplayshortskip  0pt plus1pt%
  \belowdisplayshortskip  0pt plus1pt
}
\makeatother

\begin{document}
\fancyhead{}

\title{FedMatch: Federated Learning Over Heterogeneous Question Answering Data}


\author{Jiangui Chen, Ruqing Zhang, Jiafeng Guo, Yixing Fan, and Xueqi Cheng}
\affiliation{%
  \institution{
    CAS Key Lab of Network Data Science and Technology, Institute of Computing Technology, \\ Chinese Academy of Sciences, Beijing, China\\
    University of Chinese Academy of Sciences, Beijing, China\\}
  \city{}
  \country{}
}
\email{{chenjiangui18z, zhangruqing, guojiafeng, fanyixing, cxq}@ict.ac.cn}


\begin{abstract}

Question Answering (QA), a popular and promising technique for intelligent information access, faces a dilemma about data as most other AI techniques.
On one hand, modern QA methods rely on deep learning models which are typically data-hungry. Therefore, it is expected to collect and fuse all the available QA datasets together in a common site for developing a powerful QA model.
On the other hand, real-world QA datasets are typically distributed in the form of isolated islands belonging to different parties. Due to the increasing awareness of privacy security, it is almost impossible to integrate the data scattered around, or the cost is prohibited.
A possible solution to this dilemma is a new approach known as \textit{federated learning}, which is a privacy-preserving machine learning technique over distributed datasets.
In this work, we propose to adopt federated learning for QA with the special concern on the statistical heterogeneity of the QA data.
Here the heterogeneity refers to the fact that annotated QA data are typically with non-identical and independent distribution (non-IID) and unbalanced sizes in practice. 
Traditional federated learning methods may sacrifice the accuracy of individual models under the heterogeneous situation.
To tackle this problem, we propose a novel Federated Matching framework for QA, named \textit{FedMatch}, with a backbone-patch architecture. The \textit{shared backbone} is to distill the common knowledge of all the participants while the \textit{private patch} is a compact and efficient module to retain the domain information for each participant.
To facilitate the evaluation, we build a benchmark collection based on several QA datasets from different domains to simulate the heterogeneous situation in practice. Empirical studies demonstrate that our model can achieve significant improvements against the baselines over all the datasets.

\end{abstract}

\ccsdesc[500]{Information systems~Question answering}

\keywords{Question Answering; Federated Learning; Privacy Protection}


\maketitle

\section{Introduction}

Question Answering (QA), which aims to return suitable answers in response to natural language questions issued by users \cite{yang2016anmm, laskar2020contextualized}, is a popular and crucial technique in AI. 
In recent years, QA has attracted extensive attention in both academia and industry communities due to its huge potential benefits to real-world applications, such as Amazon Alexa, Apple's Siri, Google Assistant and other intelligent information assistants.

Similar to most other AI techniques, modern QA methods face a dilemma about data. 
On one hand, deep learning models have become the major solutions ~\cite{tran2018context,tay2018multi,chen2018rnn,yoon2019compare,laskar2020contextualized} to automatically learn semantic matching between questions and answers, which requires sufficient labeled data. 
However, the labeled QA data in a single platform such as a hospital is usually limited, since data annotation is time-consuming and requires increasingly sophisticated domain knowledge. 
Therefore, it is expected to collect, fuse and use all the available QA data together in a common site for training a powerful QA model.  
On the other hand, real-world QA data usually exists in the form of isolated islands belonging to different parties. At the same time, many datasets are highly  sensitive and private, e.g., medical and legal data. 
With the increasing awareness of data security and user privacy across the world, it is almost impossible to break the barriers between data sources and integrate the data scattered around for AI processing, or the cost is prohibited.  
Therefore, how to legally solve this dilemma is a major challenge for QA researchers and practitioners today.

\begin{figure*}[t]
	\centering
	\includegraphics[scale=0.30]{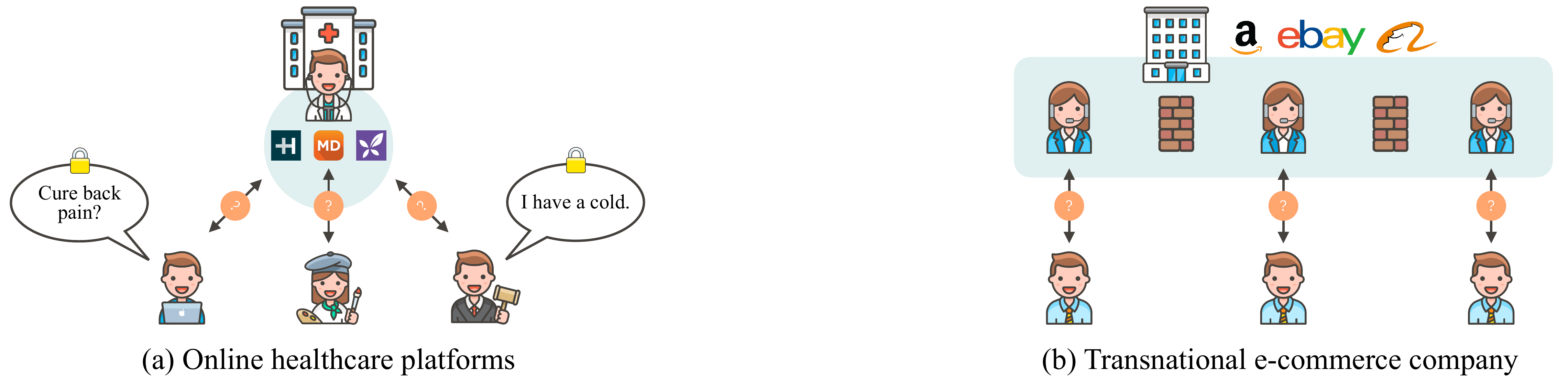}
	\caption{The statistical heterogeneity of the QA data in practice.}
	\label{fig:example}
\end{figure*}

Recently, a new privacy-preserving machine learning technique,   called \textit{federated learning}, has attracted great interest from the research community~\cite{mcmahan2017communication}.  
Specifically, the learning task is solved by a loose federation of multiple local clients which are coordinated by a central server. 
Each client has a local training dataset which is never uploaded to the server, while the server trains a global model by aggregating the local model updates. 
When the isolated data occupied by each party fails to produce an ideal model, the mechanism of federated learning makes it possible for different parties to share a united model while preventing data leakage.

Therefore, in this work, we propose to adopt federated learning for QA with the special concern on the statistical heterogeneity of the QA data. 
Here, the statistical heterogeneity refers to the fact that the annotated QA data are typically with non-identical-and-independent distribution (non-IID) and unbalanced sizes in practice.  
For example, online healthcare platforms\footnote{https://www.healthtap.com, https://www.care.com, https://portal.dxy.cn}, such as HealthTap, Care, and Ding Xiang Yuan, have been popular among patients via easing the demand for hospitals.  
As shown in Figure \ref{fig:example} (a), patients ask personalized questions related to their own disease treatment on the platform. 
The patient-doctor QA datasets have significant variances in question types, inquiry goals, as well as the case numbers.  
Another example, in international e-commerce companies\footnote{https://www.ebay.com, https://www.amazon.com, https://www.alibaba.com}, e.g., eBay, Amazon, and Alibaba, as shown in Figure \ref{fig:example} (b), users ask questions to customer services on different branch-sites.  
The customer service QA data from different branches may have remarkable gaps in language expressions, product types, as well as the data size. 
Therefore, how to model the statistical heterogeneity of the QA data becomes a critical challenge to develop an effective federated learning method for QA.

However, much of the effort has been devoted to developing federated learning methods that can better prevent privacy and integrity violations. 
Such methods may sacrifice the accuracy of individual models under the heterogeneous situation. 
To tackle such problem, we propose a novel Federated Matching framework for QA, named \textit{FedMatch}, with a backbone-patch architecture.  
Different from the original federated learning framework where all the participants share the same model, in our FedMatch method, we decompose the QA model in each participant into a shared module and a private module. 
With this framework, we are able to train a reliable unique  QA model for each participant with all participants' knowledge without exposing their raw data, which could directly work on heterogeneous data and enhance privacy protection.

Specifically, in FedMatch, the \textit{shared backbone} is used to capture the shared knowledge of different participants to empower the model training in each single participant. 
Its parameters from different participants are aggregated to update the global shared module, which is further delivered to each participant to update the local shared module.  
The \textit{private patch} is a compact and efficient module, which aims to retain the characteristics of the local data in each participant. 
We update the private patch only with the parameters computed from local stored data and exchange neither its parameters nor gradients. 
In this way, the patch can adapt to the private data distribution of each participant, and it is promising to alleviate the problem of data heterogeneity. 
Note under the client sampling setting, our framework still shows slight improvements on the performance, which again demonstrates the effectiveness of constructing a unique model for each participant.  
Besides, since the parameters of the local shared module are aggregated together, the information of labeled QA data in each participant is harder to be inferred. 
Thus, the data privacy is well-protected. 
Specifically, BERT is used as the backbone structure for storing the common parameters, while each patch is explicitly applied to each individual participant. 
We studied two types of patch architectures and four ways to insert the patch into the BERT model.

To facilitate the evaluation, we build a novel benchmark dataset FedQA\footnote{The FedQA benchmark dataset and the experimental codes are available at https://github.com/Chriskuei/FedMatch} based on several QA datasets with different sizes sourced from different domains.  
Specifically, we make use of PrivacyQA \cite{privacyqa}, BioASQ \cite{bioasq}, FiQA \cite{fiqa}, InQA \cite{inqa}, and MedQuAD \cite{medquad} datasets, to simulate the heterogeneous situation,  from law, biomedical, financial, insurance to medical.  
For evaluation, we compare with several state-of-the-art methods to verify the effectiveness of our method. 
Empirical results demonstrate that leveraging the labeled data from different QA participants in a privacy-preserving way is feasible and our proposed FedMatch framework can outperform all the baselines significantly. 
We also provide detailed analysis on the proposed framework to gain   better understanding on the learned shared and private knowledge.

\begin{figure*}[t]
	\centering
	\includegraphics[scale=0.30]{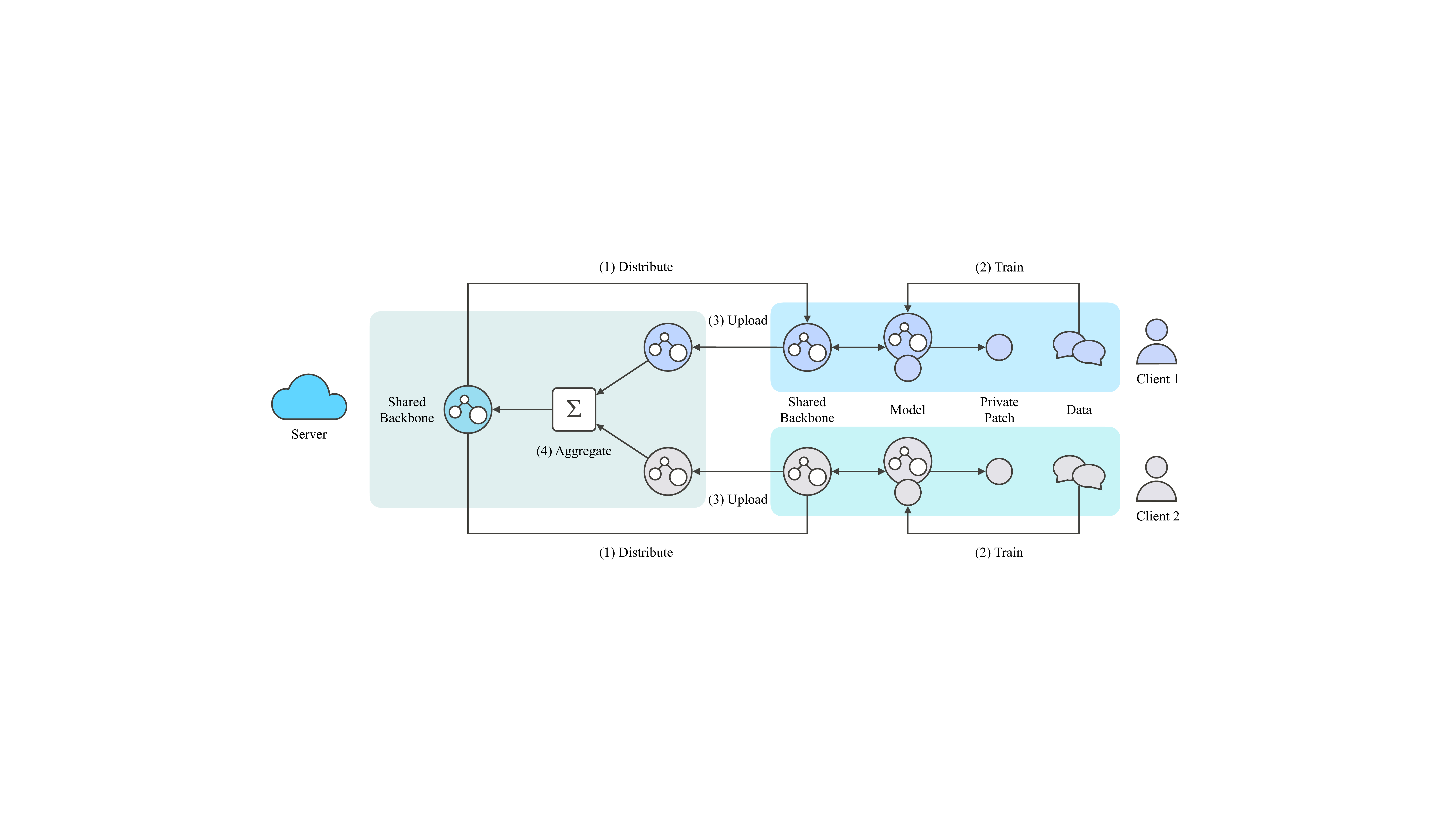}
	\caption{The overall framework of FedMatch.}
	\label{fig:framework}
\end{figure*}

\section{Related Work}

In this section, we briefly review three lines of related work, i.e., question answering, adaptation parameters and federated learning.

\subsection{Question Answering}
Question Answering (QA) aims to provide reasonable answers to users' questions. Early research works rely on different feature engineering based approaches, which extract various features from QA data to compute the matching signal~\cite{wang2007jeopardy, yih2013question, ko2007language, riezler2007statistical}. For example, ~\citet{wang2007jeopardy} proposed a statistical syntax-based model that softly aligned a question with a candidate answer and returned a score. ~\citet{yih2013question} applied rich lexical semantic information from WordNet to boost the QA matching. ~\citet{riezler2007statistical} introduced synonyms in context of the entire query by translating query terms into answer terms using a statistical machine translation model trained on QA data. Nevertheless, these feature-based approaches are usually labor-intensive, and hard to capture the semantic information between questions and answers.

With the advance of deep learning, significant improvements have been achieved on many QA tasks~\cite{shen2017inter, tran2018context, tay2018multi, chen2018rnn, yoon2019compare}. Many neural QA methods resolve lexical gaps by introducing continuous representations without preprocessing tools~\cite{mikolov2013distributed, peng2017deep}. Without loss of generality, those methods can be divided into RNN-based, CNN-based and attention-based with regard to model architecture. ~\citet{chen2018rnn} proposed a context-aligned RNN which incorporated the contextual information of the aligned words in QA data. ~\citet{shen2014learning} applied CNN to learn low-dimensional semantic vectors for questions and answers. \citet{yang2016anmm} presented an attention based neural matching model which adopted value-shared weighting scheme and incorporated question term importance.
Very recent works~\cite{yang2019end, sun2019fine, garg2020tanda, laskar2020contextualized} seek the help from pre-trained transformer-based models, e.g., BERT~\cite{devlin2018bert} and RoBERTa~\cite{liu2019roberta}. For example, \citet{laskar2020contextualized} integrated contextualized embeddings with the transformer encoder to measure the similarity of questions and answers. 
However, these methods often rely on large-scale labeled data for learning effective models, without taking into account the distributed and isolated data issues.

\subsection{Federated Learning}

Federated learning, a new privacy-preserving machine learning technique,  has attracted great interest from the research community~\cite{mcmahan2017communication}.  
Specifically, the learning task is solved by a loose federation of multiple local clients which are coordinated by a central server. 
Each client has a local training dataset which is never uploaded to the server, while the server trains a global model by aggregating the local model updates. 
When the isolated data occupied by each party fails to produce an ideal model, the mechanism of federated learning makes it possible for different parties to share a united model while preventing data leakage. 
In traditional federated learning such as FedAvg~\cite{mcmahan2017communication}, clients update all the model parameters to the server to be aggregated.
Later, there is a growing line of works demonstrating that traditional federated learning is possible to leak information about the underlying training data in unexpected ways~\cite{geiping2020inverting, melis2019exploiting, bagdasaryan2020backdoor}.
Recently, some federated learning approaches introduce the differential privacy or the robust aggregation~\cite{mcmahan2017learning, chen2019distributed, blanchard2017machine, rajput2019detox} to ensure the privacy and integrity of existing federated models.
For example, \citet{mcmahan2017learning} combined federated learning with differential privacy to formal guarantees of user-level privacy. ~\citet{chen2019distributed} replaced the average aggregation with median aggregation to prevent outliers from having much influence on the federated model.

One of the key challenges in federated learning is the data  heterogeneity problem. 
Previous studies have shown that the non-IID data distribution could  degrade the effectiveness of federated learning models~\cite{kairouz2019advances}.
In order to combat the client-drift problem caused by heterogeneous data, many approaches including FedProx~\cite{li2018federated}, SCAFFOLD~\cite{kairouz2019advances}, Mime~\cite{karimireddy2020mime} and FedNova~\cite{wang2020tackling} have been developed in recent years.
These federated optimization methods overcome the non-IID data from the aspects of regularization or control variates.
For example, \citet{li2018federated} proposed adding a proximal term to each local objective to alleviate inconsistency due to the non-IID data and heterogeneous local updates.
\citet{kairouz2019advances} introduced control variates for the server and clients, which are used to estimate the update direction of the server model and the update direction of each client.
\citet{karimireddy2020mime} used a combination of control-variates and server-level statistics at every client-update step to ensure that each local update mimics that of the centralized method run on IID data.
\citet{wang2020tackling} normalized and scaled the local updates of each client according to their number of local steps before updating the global model to ensure that the global updates are not biased.

With the increasing awareness of data security and user privacy,  federated learning has been introduced into the areas of computer vision~\cite{liang2020think, zhuang2020performance, arivazhagan2019federated} and natural language processing~\cite{liu2019roberta, singh2020pretraining, sui2020feded}.  
However, the privacy protections for the federated learning may destroy the accuracy of the federated model under the heterogeneity situations, which removes participants' main incentive to join federated learning~\cite{yu2020salvaging}. 
In this work, we propose a novel FedMatch framework with the special concern on the statistical heterogeneity.

\begin{figure*}[t]
	\centering
	\includegraphics[scale=0.58]{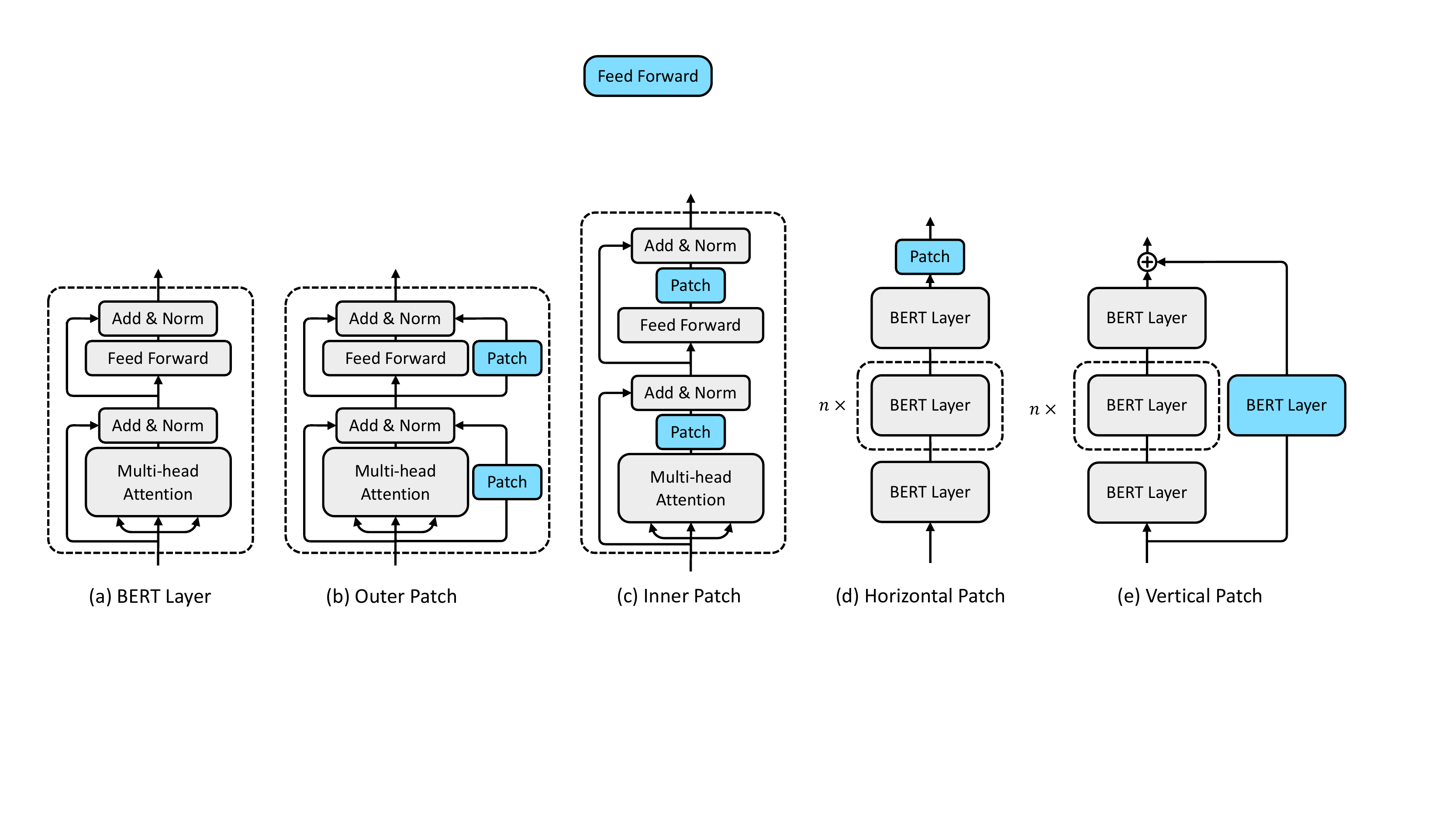}
	\caption{BERT Layer and four variants with different patch insertion ways. $l$ denotes the number of Transformer.}
	\label{fig:patch}
\end{figure*}

\section{Our Approach}

In this section, we present our proposed Federated Matching (FedMatch) framework over the heterogeneous QA data.

\subsection{Task Definition}
Question answering  devotes to assessing the relevance of a candidate answer to a given question. 
In this paper, define $T$ QA participants, each with a private QA dataset $\mathcal{D}_t \in \{ \mathcal{D}_1, \cdots, \mathcal{D}_T \}$.  
The training dataset $\mathcal{D}_t^{train}$ from participant $t$ is defined as,
\begin{equation*}
 \mathcal{D}_t^{train} = \{q_t^i, a_t^i, r_t^i\}_{i=1}^{\vert \mathcal{D}_t^{train} \vert},
\end{equation*}
where $q_t^i$, $a_t^i$ and $r_t^i \in \{0, 1\}$ denote the $i$-$th$ question, candidate answer, and matching score among $|\mathcal{D}_t^{train}|$ samples, respectively.

In the real-world situation, the annotated QA data are typically with non-identical-and-independent distribution (non-IID) and unbalanced sizes.  
Meanwhile, they are usually private and sensitive.
Therefore, the goal is to obtain a reliable unique QA model for each participant with all participants' knowledge without exposing their original data.

\subsection{Model Overview}

To tackle the federated learning for QA under heterogeneous scenario, we formulate it as a federated matching problem, i.e., to measure the relevance between the question and the answer by leveraging distributed QA datasets in a privacy-preserving way.  
We introduce a novel federated matching framework for QA, named \textit{FedMatch} for short, to solve it. 

Specifically, we consider the QA model for each participant composed of shared and private modules, which could effectively leverage the knowledge from other participants and meanwhile capture the characteristics of the local data.   
The FedMatch framework is thus designed based on this key idea which is depicted in Figure \ref{fig:framework}.

\begin{itemize}[leftmargin=*]
	\item \textbf{Common Knowledge Distillation}: The labeled data in a single participant is usually insufficient to train an accurate QA model. To alleviate the data sparsity problem, in the FedMatch framework, we propose a \textit{shared backbone}  to distill the shareable QA matching knowledge among different participants. We employ some  state-of-the-art neural matching model to assess the relevance of a candidate answer to a given question.

	\item \textbf{Domain Information Retaining}: Since the QA data stored in different participants may have different  characteristics and sizes, sharing the same model between them may not be an optimal solution. To alleviate the statistical heterogeneity, we employ  a \textit{private patch} for each participant to adapt to the specific domain  information. The patch is added to the backbone for each participant and trained only with the respective private local QA data. Consequently, the patch component assesses the participant-specific characteristics and contributes to building a unique model for each participant.  

	\item \textbf{Privacy-preserving Learning}: Sharing all training samples or model parameters among participants may make up data shortage at the sacrifice of privacy. Therefore, we propose to optimize the performance of FedMatch using the federated learning technology. We only upload parameters of the local shared module to the central server, which generally contain less privacy-sensitive information. In this way, we are able to train a reliable QA model for each participant with all participants' knowledge without exposing their original data, which enhances privacy protection.

\end{itemize}

\subsection{Shared Backbone}

The goal of the shared backbone is to learn the general and shareable knowledge for QA from multiple participants. 
In this work, we leverage the BERT \cite{devlin2018bert} as the backbone structure to measure the semantic match between questions and answers, due to its superiority in many natural language understanding tasks. 
 
Specifically, as shown in Figure \ref{fig:patch} (a), BERT's model architecture is a multi-layer bidirectional Transformer encoder \cite{vaswani2017attention} composed of a stack of identical layers, where each layer has a self-attention sub-layer and a feed-forward network sub-layer. 
We first concatenate the question and the answer to the required format, which starts with a special classification token [CLS] for the whole sequence.  
Then, with a stack of self-attention sub-layer, each token in BERT accumulates the information from both left and right context to enrich its representation.   
Finally, we apply an output softmax layer over the final hidden state of [CLS], to predict the matching score between the question and the answer. 
We now describe the self-attention and feed-forward network sub-layer in Transformer layer as follows.

\subsubsection{Self-Attention}

The Self-Attention sub-layer aims to capture global information through multi-head attention (MH) and a linear layer. 
The attention weights are derived by the dot-product similarity between transformed representations. 
Concretely, the $i$-th single attention head is,  
\begin{equation*}
	\operatorname{Attention}_i(\textbf{h}_j)=\sum_m\operatorname{softmax}(\frac{W_i^q\textbf{h}_j\cdot W_i^k\textbf{h}_m}{\sqrt{d/n}})W_i^v\textbf{h}_m,
\end{equation*}
where $\textbf{h}_j$ denotes a $d$ dimensional hidden vector of the $j$-th sequence token. 
$W_i^q, W_i^k, W_i^v$ are learned matrices of size $d/n \times d$. 
Finally, the outputs of the $n$ attention heads are concatenated together and passed to a linear transformation, 
\begin{equation*}
	\operatorname{MH}(\textbf{h})=\operatorname{Concat}(\operatorname{Attention}_1(\textbf{h}), \dots, \operatorname{Attention}_n(\textbf{h}))W^o,
\end{equation*}
where the learned matric $W^o \in \mathbb{R}^{d\times d}$. 
The outputs are further passed to a residual connection followed by layer normalization \cite{ba2016layer}. 
We denote this process as $\operatorname{SA}(\cdot)$,
\begin{equation*}
	\operatorname{SA}(\textbf{h})=\operatorname{LN}(\operatorname{MH}(\textbf{h})+\textbf{h}),
\end{equation*}
where $\operatorname{LN}(\cdot)$ is layer normalization. 

\subsubsection{Feed-forward Network}

The feed-forward network is a position-wise fully connected feed-forward network (FFN), which is applied to each position separately and identically, i.e., 
\begin{equation*}
	\operatorname{FFN}(\textbf{h})=W_2f(W_1\textbf{h}+b_1)+b_2,
\end{equation*}
where $f(\cdot)$ is an activation function~\cite{he2016deep}. 
$W_1, W_2, b_1,$ and $b_2$ are learned matrices. 

Putting this together, a BERT layer $\operatorname{BL}(\cdot)$ is a layer-norm (LN) applied to the output of FFN layer, with a residual connection, 
\begin{equation*}
	\operatorname{BL}(\textbf{h})=\operatorname{LN}(\operatorname{FFN}(\operatorname{SA}(\textbf{h}))+\operatorname{SA}(\textbf{h})).
\end{equation*}

\subsection{Private Patch}

To model the domain information for each participant, inspired by ~\cite{houlsby2019parameter}, we introduce a compact and efficient module, i.e., the private patch, for each participant based on the local data. 
The private patch is responsible for adapting exiting BERT representation in the shared backbone to the specific domain. 
We now describe different ways of the patch insertion and the patch structure.

\subsubsection{Patch Insertion}

Here, we introduce where to insert the patch into the BERT model. 
Based on the previous introduction, each BERT layer contains a self-attention and a feed forward layer. 
As shown in Figure~\ref{fig:patch}, we explore four positions to insert patch, including inner, outer, horizontal, and vertical. 
Specifically, we add the patch to the two sub-layers in each BERT layer in the inner and outer fashion, while we add the patch to the BERT in the vertical and horizontal fashion.

\begin{itemize}[leftmargin=*]
\item \textbf{Inner}. In inner fashion, as shown in  Figure~\ref{fig:patch}(b), we put a patch following the self attention layer, another following the feed forward layer. 
In this way, the output of the self-attention layer and the BERT layer are as follows:
\begin{equation*}
	\operatorname{SA}(\textbf{h})=\operatorname{LN}(Patch(\operatorname{MH}(\textbf{h}))+\textbf{h}),
\end{equation*}
\begin{equation*}
	\operatorname{BL}(\textbf{h})=\operatorname{LN}(Patch(\operatorname{FFN}(\operatorname{SA}(\textbf{h}))) +\operatorname{SA}(\textbf{h})),
\end{equation*}
where $Patch(\cdot)$ denotes the patch layer.

\item \textbf{Outer}. In outer fashion, as shown in Figure~\ref{fig:patch}(c), we add two patches parallel with the self-attention layer and the feed forward layer respectively. 
In this way, the output of the self-attention layer and the BERT layer are as follows:
\begin{equation*}
	\operatorname{SA}(\textbf{h})=\operatorname{LN}(\operatorname{MH}(\textbf{h})+\textbf{h}+Patch(\textbf{h})),
\end{equation*}
\begin{equation*}
	\operatorname{BL}(\textbf{h})=\operatorname{LN}(\operatorname{FFN}(\operatorname{SA}(\textbf{h}))+ Patch(\operatorname{SA}(\textbf{h}))+ \operatorname{SA}(\textbf{h})).
\end{equation*}

\item \textbf{Vertical}. In vertical fashion, as shown in Figure~\ref{fig:patch}(d), we put a patch following the topmost layer in BERT. 
In this way, the output of the BERT is defined as, 
\begin{equation*}
	\operatorname{BL}_{topmost}(\textbf{h})=Patch(\operatorname{BL}_{topmost}(\textbf{h})),
\end{equation*}
where $\operatorname{BL}_{topmost}(\cdot)$ denotes the output of BERT's topmost layer.

\item \textbf{Horizontal}. In horizontal fashion, as shown in Figure~\ref{fig:patch}(e), we add a patch parallel with each BERT layer.
In this way, the output of the BERT layer is defined as,  
\begin{equation*}
	\operatorname{BL}(\textbf{h})=Patch(\textbf{h}) + \operatorname{BL}(\textbf{h}).
\end{equation*}

\end{itemize}

\begin{figure}[t]
	\centering
	\includegraphics[scale=0.58]{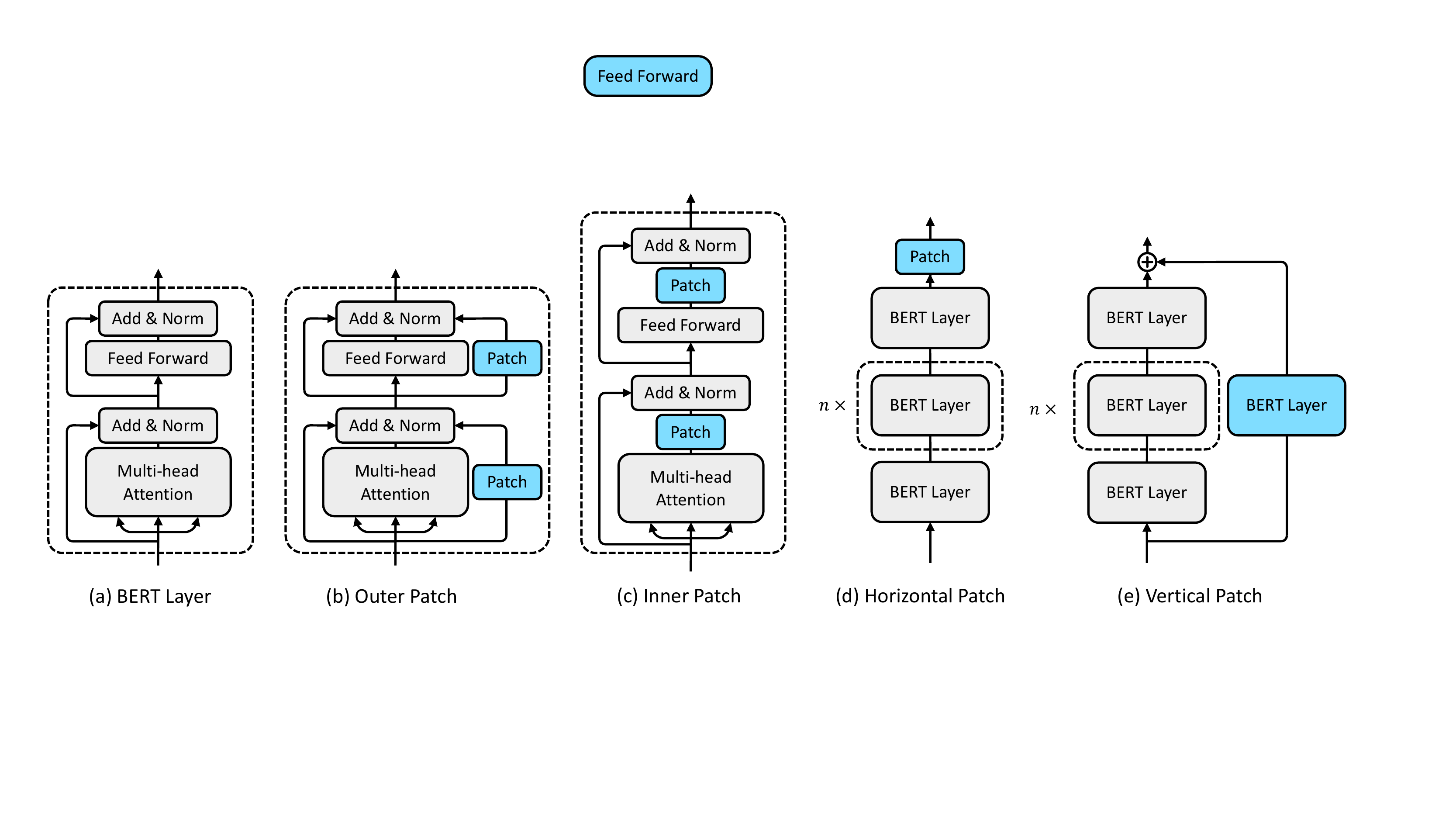}
	\caption{Two types of patch architectures.}
	\label{fig:patch_architectures}
\end{figure}

\subsubsection{Patch Structure}

Here, we introduce how to design the patch structure. 
The patch structure for each participant is defined as,
\begin{equation*}
	Patch(\textbf{h})=V^Dg(V^E\textbf{h}),
\end{equation*} 
where $V^E\in \mathbb{R}^{d_s\times d}$ and $V^D\in \mathbb{R}^{d\times d_s} (d_s < d)$. $g(\cdot)$ is an arbitrary function.
Existing works have explored many forms of $g(\cdot)$, achieving an effective model capacity with the same number of parameters in multi-task learning and continual learning~\cite{houlsby2019parameter, cooper2019bert, collobert2008unified}. 

Here, as shown in Figure~\ref{fig:patch_architectures}, we investigate two forms of $g(\cdot)$ to efficiently enlarge the model capacity, including Projected Attention Layer and Low-rank Layer.
\begin{itemize}[leftmargin=*]
	\item \textbf{Projected Attention Layer (PAL)}~\cite{cooper2019bert}. The $g(\cdot)$ is defined as a multi-head attention layer. The intuition is that different participants need different interactions between token representations. 
	\item \textbf{Low-Rank Layer}~\cite{cooper2019bert}. The $g(\cdot)$ is a low-rank linear transformation, i.e., a standard feed-forward network.   
\end{itemize}


\subsection{Federated Training}\label{sec:training}

\begin{table*}[t]
\caption{Overall statistics of FedQA benchmark dataset. \#Questions: the number of questions, \#Answers: the number of answers, \#Avg QL: the average length of questions, \#Avg AL: the average length of answers, \%PosRate: the rate of positive labels.}
\label{tab:benchmark}
\renewcommand{\arraystretch}{1.1}
   \setlength\tabcolsep{8pt}
\begin{tabular}{lcccccc}
\toprule \hline
\textbf{Dataset} & \textbf{Domain} & \textbf{\#Questions} & \textbf{\#Answers} & \textbf{\#Avg QL} & \textbf{\#Avg AL} & \textbf{\%PosRate} \\
\midrule
\textbf{PrivacyQA} & \textsc{Law}  & 1,750 & 4,947 & 8.46  & 139.62 & 14.29 \\
\textbf{BioASQ} & \textsc{Biomedical} & 2,740 & 12,815 & 10.5 & 36.0 & 21.87 \\
\textbf{FiQA} & \textsc{Financial} & 6,648 & 26,016 & 12.4 & 202.0 & 32.15 \\
\textbf{INQA} & \textsc{Insurance} & 1,309 & 27,413 & 7.2  & 92.3 & 25.67 \\
\textbf{MedQuAD} & \textsc{Medical} & 380 & 2,396 & 19.7 & 469.7 & 37.22\\
\hline \bottomrule
\end{tabular}
\end{table*}

To protect participants' privacy, we utilize federated learning to train the QA model with a backbone-patch architecture, over data from different participants.  
In FedMatch, as shown in Figure \ref{fig:framework}, the central server coordinates multiple clients for patch updating and backbone sharing. 
Specifically, the clients here are different QA participants, and train their models with privately stored data. 
The server first initializes the parameters $\theta$ of the shared backbone randomly, and the training phase includes the following steps:

\begin{enumerate}[leftmargin=*]
\item The server distributes the parameters $\theta$ of the global shared backbone to each client for the next-round model training. 
\item Each client adds a private patch to the shared backbone, and trains their local models based on privately stored data. Formally, for each client $t$, let $\theta_t$ denote the parameters of the local shared backbone and $\beta_t$ denote the parameters of the private patch. The loss function for each client $t$ is a pairwise ranking loss over the training dataset $\mathcal{D}_t^{train}$, i.e., 
 \begin{equation*}
\mathcal{L}_{t}(q_t, a_t^{+}, a_t^{-};\Theta_t)=\max(0, 1-f(q_t, a_t^{+})+f(q_t, a_t^{-}))
\end{equation*}
where $f(\cdot)$ denotes a QA matching model. $a_t^{+}$ and $a_t^{-}$ denotes a relevant answer and a negative answer with respect to the question $q_t$ respectively. $\Theta_t = (\theta_t, \beta_t)$ denotes all parameters of the local client model. Specifically, $\theta_t$  is initialized using the shared model's parameters $\theta$ and $\beta_t$ is randomly initialized.

\item For every training epoch, the goal is to minimize a global loss   over all $T$ distributed clients, i.e.,  
\begin{equation*}
  minimize_{\{\Theta_1, \cdots, \Theta_T\}}\mathcal{L}(\Theta_1, \cdots, \Theta_T). 
\end{equation*}
After each training epoch, each client updates the parameters $\theta_t$ of the local shared backbone to the server. 

\item The server monitors each client for parameter aggregation and performs global backbone updating once it has collected parameters from all clients. 
 Formally, given the parameters from $T$ clients, we update the parameters of the globally shared backbone stored on the central server, 
\begin{equation*}
	\theta=\frac{1}{T}\sum_{t=1}^T\theta_t.
\end{equation*}   
	
\end{enumerate}

The process described above is repeated iteratively until the entire model converges.

\section{Experiments}

In this section, we conduct experiments to verify the effectiveness of our proposed model.

\subsection{Benchmark Construction}

In order to facilitate the study of federated learning for QA,  we build a new benchmark dataset FedQA based on several public QA collections.

\begin{itemize}[leftmargin=*]
 \item \textbf{PrivacyQA}~\cite{privacyqa} is a corpus about the privacy policies of mobile applications representing different categories. Crowd workers ask privacy questions about a given mobile application. And then the authors recruit seven experts with legal training to construct answers to questions. 
 \item \textbf{BioASQ}~\cite{bioasq} is a competition on biomedical semantic indexing and QA. Biomedical workers are allowed to express their information needs, and then concise answers are returned by combining information from multiple sources of different kinds. 
 \item \textbf{FiQA}~\cite{fiqa} is created for WWW'18 financial opinion mining and QA challenge. We leverage the data of Task 2, i.e., Opinion-based QA over financial data. Questions are answered based on a corpus of documents from different financial data sources.
 \item \textbf{InQA}~\cite{inqa} collects the question and answer pairs from the insurance domain, driven by the intense scientific and commercial interest in this domain. The questions are collected from real-world users, and the answers are composed by professionals with deep domain knowledge.
 \item \textbf{MedQuAD}~\cite{medquad} is a collection of QA pairs from the medical domain, constructed from 12 trusted websites. The collection includes 16 types about Diseases, 20 types about Drugs and 1 type for the other named entities. 
\end{itemize}

Table \ref{tab:benchmark} shows the overall statistics of our FedQA  benchmark dataset. 
We take these five collections as our whole QA datasets, since (1) These collections are publicly available; 
(2) The contexts in these collections are different from each other and it is reasonable to distinguish one domain from another domain. 
Besides, these collections have significant variances in the numbers of QA pairs. 
For example, the number of financial questions in FiQA is about 17 times as that of medical questions in MedQuAD.  
In this way, we can obtain a 5-domain dataset to mimic the statistical heterogeneity, which makes the federated learning for QA more challenging and closer to the real-world situation.

\subsection{Experimental Settings}

To evaluate the performance of our method, we conduct experiments on our FedQA benchmark dataset. 
For pre-processing, all the words in answers and questions are white-space tokenized and lower-cased. 
We leverage Elasticsearch\footnote{https://www.elastic.co} to index all the answers in FedQA using BM25 \cite{robertson2009probabilistic}. 
Since there are no negative samples in the training set, we take the top 5 retrieved results which are not the ground-truth answers as  the negative samples.

We implement our model in PyTorch\footnote{https://pytorch.org} based on Transformers library\footnote{https://github.com/huggingface/transformers}. 
We optimize the model using Adam \cite{kingma2014adam} with the warmup technique, where the learning rate increases over the first 10\% of batches, and then decays linearly to zero. 
The learning rate for each QA collection in FedQA is set to $2e^{-5}$, and the batch size is \{32, 32, 32, 32, 12\} for PrivacyQA, BioASQ, FiQA, INQA, and MedQuAD respectively.
All runs are trained on a Tesla 32G V100 GPU. 
The dimension of the transformation matrix in Low-Rank layer is 128. 
We use the base-uncased version of BERT. 
All hyper-parameters of our model are also tuned using the development set.
Due to the datasets limitation, we only consider the cross-silo setting. Specifically, each QA dataset is regarded as one participant and all participants are trained in a communication round.

By combining four ways of patch insertion (i.e., inner, outer, vertical, and horizontal) and two patch structures (i.e., PAL and Low-Rank), we obtain eight types of $FedMatch$ denoted as $FedMatch_{I+PAL}$, $FedMatch_{I+LR}$, $FedMatch_{O+PAL}$, $FedMatch_{O+LR}$, $FedMatch_{V+PAL}$, $FedMatch_{V+LR}$, $FedMatch_{H+PAL}$, and $FedMatch_{H+LR}$. 

\begin{table*}[t]
\caption{Model analysis of our FedMatch model under the MAP and MRR metric.}
\label{tab:ablation}
\renewcommand{\arraystretch}{1.2}
\begin{tabular}{lcccccccccccc}
\toprule \toprule
& \multicolumn{2}{c}{\textbf{PrivacyQA}} & \multicolumn{2}{c}{\textbf{BioASQ}} & \multicolumn{2}{c}{\textbf{FiQA}}  & \multicolumn{2}{c}{\textbf{INQA}} & \multicolumn{2}{c}{\textbf{MedQuAD}} & \multicolumn{2}{c}{\textbf{Overall}}           \\
\cmidrule(lr){2-3} \cmidrule(lr){4-5} \cmidrule(lr){6-7} \cmidrule(lr){8-9} \cmidrule(lr){10-11} \cmidrule(lr){12-13}
\textbf{Method} & \textbf{MAP} & \textbf{MRR} & \textbf{MAP} & \textbf{MRR} & \textbf{MAP} & \textbf{MRR} & \textbf{MAP} & \textbf{MRR} & \textbf{MAP} & \textbf{MRR} & \textbf{MAP} & \textbf{MRR} \\
\midrule
$\textbf{FedMatch}_{I+PAL}$  & 0.6816 & 0.6816 & 0.8640 & 0.8793 & 0.7959 & 0.8531 & 0.8483 & 0.8802 & 0.8415 & 0.9134 & 0.8063 & 0.8415 \\
$\textbf{FedMatch}_{I+LR}$   & 0.6849 & 0.6849 & 0.8622 & 0.8808 & 0.7935 & 0.8498 & 0.8551 & 0.8877 & \textbf{0.8443} & \textbf{0.9239} & 0.8080 & 0.8454 \\
\midrule
$\textbf{FedMatch}_{O+PAL}$  & 0.6826 & 0.6826 & 0.8577 & 0.8728 & 0.7941 & 0.8518 & 0.8619 & 0.8935 & 0.8310 & 0.9116 & 0.8055          & 0.8425 \\
$\textbf{FedMatch}_{O+LR}$   & 0.6987 & 0.6987 & 0.8600 & 0.8762 & 0.7947 & 0.8488 & 0.8621 & 0.8971 & 0.8415 & 0.9113 & 0.8114 & 0.8464 \\
\midrule
$\textbf{FedMatch}_{V+PAL}$  & 0.7036 & 0.7036 & 0.8421 & 0.8580 & 0.7926 & 0.8494 & 0.8614 & 0.8944 & 0.8258 & 0.9055 & 0.8051 & 0.8422 \\
$\textbf{FedMatch}_{V+LR}$   & 0.7036 & 0.7036 & \textbf{0.8673} & \textbf{0.8837} & \textbf{0.7958} & \textbf{0.8549} & 0.8564 & 0.8920 & 0.8385 & 0.9215 & 0.8123 & 0.8511 \\
\midrule
$\textbf{FedMatch}_{H+PAL}$  & 0.7200 & 0.7200 & 0.8531 & 0.8699 & 0.7904 & 0.8494 & 0.8632 & 0.9080 & 0.8414 & 0.9186 & 0.8136 & 0.8532 \\
$\textbf{FedMatch}_{H+LR}$   & \textbf{0.7251} & \textbf{0.7251} & 0.8591 & 0.8734 & 0.8047 & 0.8503 & \textbf{0.8641} & \textbf{0.9015} & 0.8435 & 0.9214 & \textbf{0.8193} & \textbf{0.8543} \\
\bottomrule \bottomrule
\end{tabular}
\end{table*}

\subsection{Evaluation Metrics}

For evaluation, we employ the overall performance on all test datasets from $T$ different participants,
\begin{equation*}
 \frac{1}{T}\sum_{t=1}^T\frac{1}{\vert \mathcal{D}_t^{test} \vert} \cdot \sum_{j=1}^{\vert \mathcal{D}_t^{test} \vert} g(r_t^j, f^{\ast}(\{q_t^j, a_t^j\};\Theta_t),
\end{equation*}
where $\mathcal{D}_t^{test}$ denotes the test dataset from participant  $t$, and $q_t^j$, $a_t^j$ and $r_t^j \in \{0, 1\}$ denote the $j$-$th$ question, candidate answer, and matching score among $|\mathcal{D}_t^{test}|$ test samples, respectively. $f^{\ast}(\cdot)$ denotes the learned QA matching model for each participant, and $g(\cdot)$ denotes the evaluation metric for QA.  
Following \cite{yang2016anmm, laskar2020contextualized}, we leverage two widely used metrics, i.e., MAP and MRR, as the implementation of $g(\cdot)$.



The collections from five domains in FedQA are distributed in five clients. 
Inspired by \cite{arivazhagan2019federated}, we report the performance of each client.  
We also show the \textbf{overall} performance of all the clients via computing the average of evaluation metrics in each domain.

\subsection{Baselines}

We adopt three types of baseline methods for comparison, including individual methods, privacy enhanced methods, and conventional federated learning methods.

\subsubsection{Individual Methods} We first compare our methods with several QA models without the use of federated learning. 
\begin{itemize}[leftmargin=*]

\item \textbf{RE2}~\cite{re2} highlights three key features, namely previously aligned features, original point-wise features, and contextual features for inter-sequence alignment.   
\item \textbf{ESIM}~\cite{esim} uses Bi-LSTM to encode texts and applies the attention and fusion layer over the representations to obtain the label. 
\item \textbf{BERT}~\cite{devlin2018bert} denotes that BERT$_{base}$ is fine-tuned locally on QA data in each individual participant. 

\end{itemize}

\subsubsection{Traditional Privacy Enhanced Methods}

We also apply one traditional privacy enhanced model.

\begin{itemize}[leftmargin=*]
	\item \textbf{CoverQuery}~\cite{ahmad2016topic} generates several noisy queries from unrelated topics to hide the original data, which is widely used in personalized web search. 
\end{itemize}


\subsubsection{Conventional Federated Learning Methods}

In the original design, the federated learning method is created by repeatedly averaging model updates from small subsets of participants. 
\begin{itemize}[leftmargin=*]
	\item \textbf{FedAvg}~\cite{mcmahan2017communication} combines local stochastic gradient descent on each client with a server that performs model averaging.
	\item \textbf{LG-FedAvg}~\cite{liang2020think} denotes the local global federated averaging, where the global model only operates on local representations to reduce the number of communicated parameters. 
	\item \textbf{FedPer}~\cite{arivazhagan2019federated} comprises of the base layers being trained by federated averaging and personalization MLP layers being trained only from local data. 
\end{itemize}

\subsection{Model Analysis}
We first analyze our models using different ways of patch insertion (i.e., inner, outer, vertical, and horizontal) and different patch structures (i.e., PAL and Low-Rank). 
As shown in Table~\ref{tab:ablation}, we have the following observations: 
(1) \textit{FedMatch} with the \textit{PAL} patch structure performs worse than that with the \textit{low-rank} patch structure in terms of the overall model performance. 
The results indicate that the simple low-rank transformation has greater task-specific representational capacity.    
(2) The \textit{vertical} and \textit{horizontal} ways of patch insertion are more effective than the \textit{inner} and \textit{outer} ways.
For example, the relative improvement of $FedMatch_{H+PAL}$ over $FedMatch_{O+PAL}$ is about 1.27\% in terms of MRR on the overall performance. 
The reason might be that retaining the global domain information via inserting patch at the BERT-level more closely resembles the QA matching process. 
(3) $FedMatch_{H+LR}$ achieves the best performance in terms of the overall model performance, showing the effectiveness of inserting the low-rank patch structure into the backbone in the horizontal fashion.

\begin{table*}
\caption{Comparisons between our FedMatch and the baselines under the  MAP and MRR metric. Two-tailed t-tests demonstrate the improvements of FedMatch over the representative method BERT are statistically significant ($^\ddag$ indicates p-value < 0.05).}
\label{tab:results}
\setlength{\tabcolsep}{1.4mm}
\renewcommand{\arraystretch}{1.1}
\begin{tabular}{lcccccccccccc}
\toprule \toprule
& \multicolumn{2}{c}{\textbf{PrivacyQA}} & \multicolumn{2}{c}{\textbf{BioASQ}} & \multicolumn{2}{c}{\textbf{FiQA}}  & \multicolumn{2}{c}{\textbf{INQA}} & \multicolumn{2}{c}{\textbf{MedQuAD}} & \multicolumn{2}{c}{\textbf{Overall}}           \\
\cmidrule(lr){2-3} \cmidrule(lr){4-5} \cmidrule(lr){6-7} \cmidrule(lr){8-9} \cmidrule(lr){10-11} \cmidrule(lr){12-13}
\textbf{Method} & \textbf{MAP} & \textbf{MRR} & \textbf{MAP} & \textbf{MRR} & \textbf{MAP} & \textbf{MRR} & \textbf{MAP} & \textbf{MRR} & \textbf{MAP} & \textbf{MRR} & \textbf{MAP} & \textbf{MRR} \\
\midrule
\textbf{ESIM} & 0.6809 & 0.6809 & 0.7378 & 0.7564 & 0.6736 & 0.7378 & 0.8533 & 0.8884 & 0.7629 & 0.8396 & 0.7417 & 0.7806 \\
\textbf{RE2}  & 0.6839 & 0.6839 & 0.7557 & 0.7697 & 0.7263 & 0.7911 & 0.8656 & 0.8990 & 0.7460 & 0.8262 & 0.7555 & 0.7940\\
\textbf{BERT}  & 0.6912 & 0.6912 & 0.8580 & 0.8650 & 0.8042 & 0.8430 & 0.8059 & 0.8520 & 0.8093 & 0.8888 & 0.7937 & 0.8278 \\
\midrule
\textbf{CoverQuery} & 0.6772 & 0.6772 & 0.8458 & 0.8627 & 0.7846 & 0.8370 & 0.8262 & 0.8558 & 0.8144 & 0.8923 & 0.7896 & 0.8250 \\
\midrule
\textbf{LG-FedAvg} & 0.7028 & 0.7028 & 0.8546 & 0.8721 & 0.7858 & 0.8435 & 0.7977 & 0.8435 &  0.8208 & 0.9076 & 0.7923 & 0.8339\\ 
\textbf{FedPer} & 0.6962 & 0.6962 & 0.8449 & 0.8625 & 0.7839 & 0.8389 & 0.8219 & 0.8574 & 0.8304 & 0.9210 & 0.7955 & 0.8352 \\
\textbf{FedAvg} & 0.6746 & 0.6746 & 0.8575 & 0.8720 & 0.7798 & 0.8359 & 0.8419 & 0.8850 & 0.8433 & 0.9177 & 0.8006 & 0.8370 \\
$\textbf{FedMatch}$   & 0.7251$^\ddag$ & 0.7251$^\ddag$ & \textbf{0.8591} & 0.8734$^\ddag$ & \textbf{0.8047} & 0.8503$^\ddag$ & \textbf{0.8641}$^\ddag$ & \textbf{0.9015}$^\ddag$ & 0.8435$^\ddag$ & 0.9214$^\ddag$ & \textbf{0.8193}$^\ddag$ & 0.8543$^\ddag$ \\
\midrule
$\textbf{FedAvg}_{CS}$ & 0.6709 & 0.6709 & 0.8506 & 0.8691 & 0.7871 & 0.8439 & 0.8129 & 0.8622 & 0.8090 & 0.8854 & 0.7861 & 0.8263\\
$\textbf{FedMatch}_{CS}$ & \textbf{0.7309}$^\ddag$ & \textbf{0.7309}$^\ddag$ & 0.8573 & \textbf{0.8741}$^\ddag$ & 0.7886 & \textbf{0.8516}$^\ddag$ & 0.8630$^\ddag$ & 0.8944$^\ddag$ & \textbf{0.8506}$^\ddag$ & \textbf{0.9276}$^\ddag$ & 0.8181$^\ddag$ & \textbf{0.8557}$^\ddag$ \\
\bottomrule \bottomrule
\end{tabular}
\end{table*}

\subsection{Baseline Comparison}

The performance comparisons between our model and the baselines are shown in Table~\ref{tab:results}. 
We can observe that: 
(1) The conventional federated learning method \textit{FedAvg}  outperforms individual models (i.e., \textit{RE2}, \textit{ESIM} and \textit{BERT}) in terms of the overall MAP and MRR performance. 
The results show that compared with training model on the data of a single participant, the federated learning method could train more accurate QA model by leveraging the useful information from multiple participants. 
(2) Individual models could outperform \textit{FedAvg}, \textit{LG-FedAvg} and \textit{FedPer} on some domains. 
For example, as compared with \textit{FedAvg}, the relative improvement of \textit{BERT} over the FiQA set in terms of MAP is about 3.13\%.  
The reason might be that \textit{FedAvg} trains a single model for all clients, making it difficult to model the statistical heterogeneity of the FedQA benchmark. 
(3) The performance of \textit{CoverQuery} has a significant drop as compared with original federated learning frameworks (i.e., \textit{FedAvg}, \textit{LG-FedAvg} and \textit{FedPer}).  
The results indicate that federated learning is more effective than the traditional privacy enhanced methods while providing more privacy guarantee.
(4) Our \textit{FedMatch} model achieves the best performance.  
The results validate the effectiveness of our strategy in decomposing the QA model into a shared backbone to learn the general knowledge from multiple clients, and a private patch to capture the local data characteristics.

\subsection{Impact of Client Sampling}

Some recent works have shown that once the client-sampling setting is considered, the non-IID data could significantly affect the performance~\cite{karimireddy2020mime}. 
Specifically, we further analyze the performance of \textit{FedMatch} and the representative federated learning baseline \textit{FedAvg} under client sampling.  
Here, we randomly sample 2 clients in each communication round. 
We denote \textit{FedMatch} and \textit{FedAvg} under client sampling as \textit{FedMatch$_{CS}$} and \textit{FedAvg$_{CS}$} respectively.  
As shown in Table~\ref{tab:results}, we can find that: 
(1) \textit{FedAvg$_{CS}$}  performs worse than \textit{FedAvg}. It again implies that the non-IID distribution is a critical challenge for standard federated learning method \textit{FedAvg} which trains a single global model for all clients. 
(2) \textit{FedMatch$_{CS}$} shows slight improvements over \textit{FedMatch}. 
The results imply the responsibility of the unique model designed  for each client in our framework, which could adapt to the privately data distribution. 
In this way, it is promising to alleviate the problem of overall  non-IID data distribution.

\subsection{Impact of Shared Backbone Size}

Since BERT is used as the backbone structure for storing the common parameters, we would like to study the effect of different sizes of BERT's shared layers on the QA performance. 
There are $12$ layers in BERT$_{base}$ and we successively make the top layers private.  We compare the performance of $FedMatch_{H+LR}$ using different numbers of BERT's shared layers, varying in the range of $[12,0]$, where $12$ denotes that all the layers are shared and $0$ denotes that BERT is totally private.

\begin{figure}[h]
	\centering
	\includegraphics[scale=0.66]{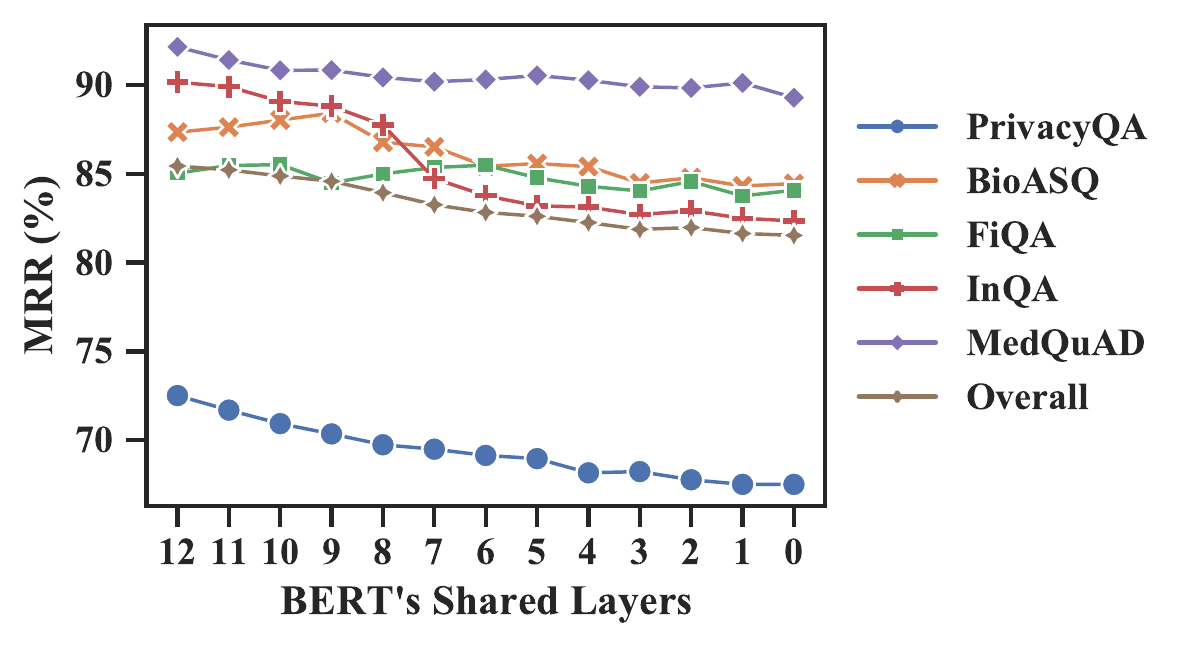}
	\caption{Performance comparison of the FedMatch$_{H+LR}$ method with different sizes of BERT's shared layers.}
	\label{fig:shared}
\end{figure}

Figure~\ref{fig:shared} shows the MRR performance over different sets with the decrease of BERT's shared layers. 
We can see that:  
(1) The overall performance of all the collections (brown color) decreases with the decrease of shared layers. 
It indicates that most participants could gain more benefit from more shared knowledge of all the participants. 
(2) An interesting phenomenon is that the performance over the FiQA is more robust than that over other datasets. 
The reason might be that the FiQA has enough high-quality financial QA data for training a powerful model, resulting the little affect by the shared knowledge.  
(3) For the BioASQ and the FiQA, the performance is not optimal if all the participants share all the 12 layers. 
The reason might be that the data on different participants usually has different characteristics, which can not be captured if we constraint different  participants to share the entire BERT.

\subsection{Impact of Private Patch Size}

\begin{figure}[h]
	\centering
	\includegraphics[scale=0.66]{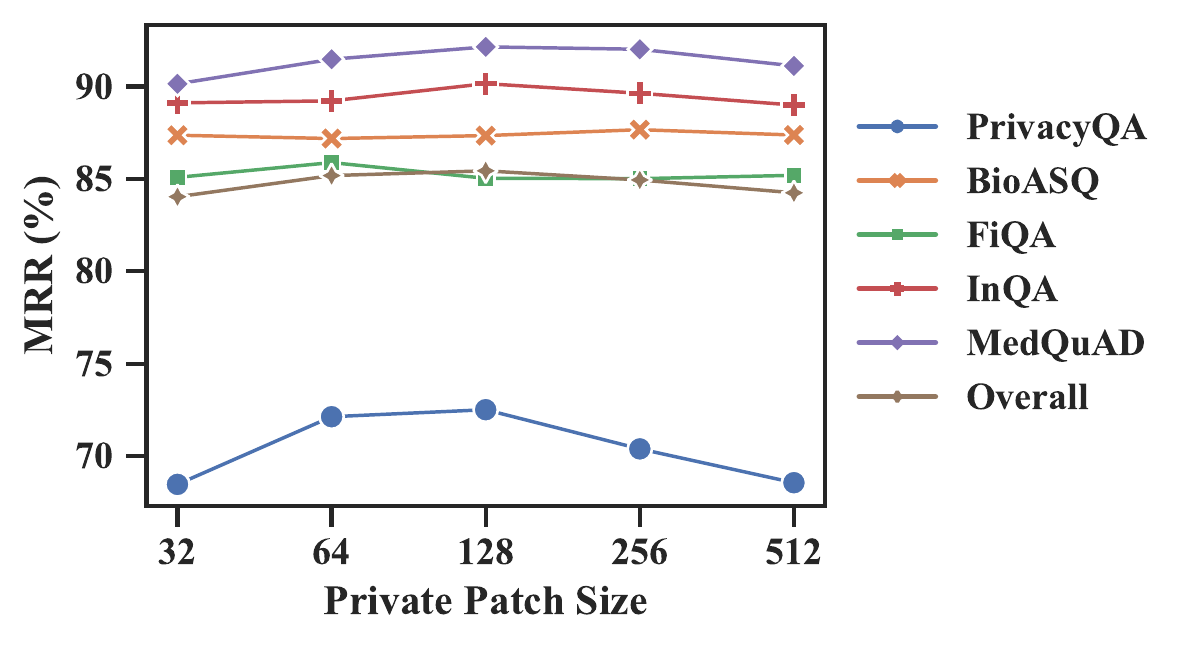}
	\caption{Performance comparison of the FedMatch$_{H+LR}$ method over different sizes of private patch.}
	\label{fig:private}
\end{figure}

The private patch size, i.e., the dimension of projection space, is a hyper-parameter in our proposed FedMatch model. 
Smaller patch consumes fewer parameters while the performance may decrease. 
Larger patch may improve the performance while the parameters could be large. 
Here, we test the performance over different sizes of the low-rank layer in FedMatch$_{H+LR}$, and vary the size in $\{32, 64, 128, 256, 512\}$. 
As shown in Figure \ref{fig:private}, we can find that when the patch size exceeds some threshold, the $FedMatch_{H+LR}$ performs worse as the patch becomes bigger. 
A possible reason is the overfitting of the patch for the participant. 
For example, by introducing less than 128 or more than 128 patch sizes, our private structure over the PrivacyQA data tends to capture insufficient information or noisy information that may hurt the matching performance.
Therefore, it is necessary to achieve a trade-off between the patch size and the performance.  

\subsection{Impact of Global Aggregation Frequency}

\begin{figure}[h]
	\centering
	\includegraphics[scale=0.66]{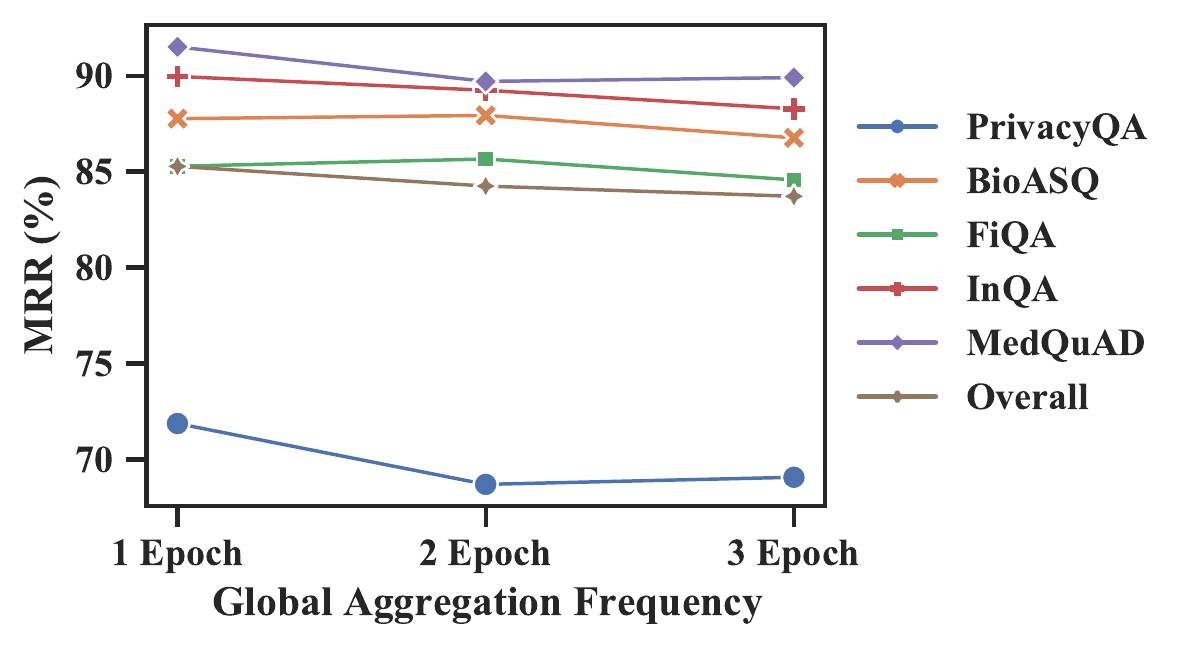}
	\caption{Performance comparison of the FedMatch$_{H+LR}$ method over different frequencies of global aggregation.}
	\label{fig:frequency}
\end{figure}

In our $FedMatch$ model, the server first aggregates received model updates from multiple participants after each epoch. Then, the server updates the global model and distributes the new model to each client for next-round training. 
Here we analyze the effect of different frequencies of the global aggregation, i.e., 1, 2, and 3 epochs. 
As shown in Figure~\ref{fig:frequency}, we can find that: 
(1) For most datasets (i.e., PrivacyQA, InQA, and MedQuAD), $FedMatch_{H+LR}$ can achieve the best performance if the global aggregation is executed every training epoch. 
For the BioASQ and the FiQA, $FedMatch_{H+LR}$ achieves the best performance if the global aggregation is executed every two training epochs, which improves the results a little over every training epoch. 
(2) $FedMatch_{H+LR}$ on large datasets such as the FiQA and the BioASQ, are more robust with respect to the aggregation frequency. 
(3) The performance of every two epoch of aggregation over the PrivacyQA has a significant drop as compared that of every epoch. 
The reason might be that longer training on local data could result in more loss of generalized knowledge. 

\begin{figure}[t]
 \centering
 \subfigure[MAP]{
 \includegraphics[scale=0.50]{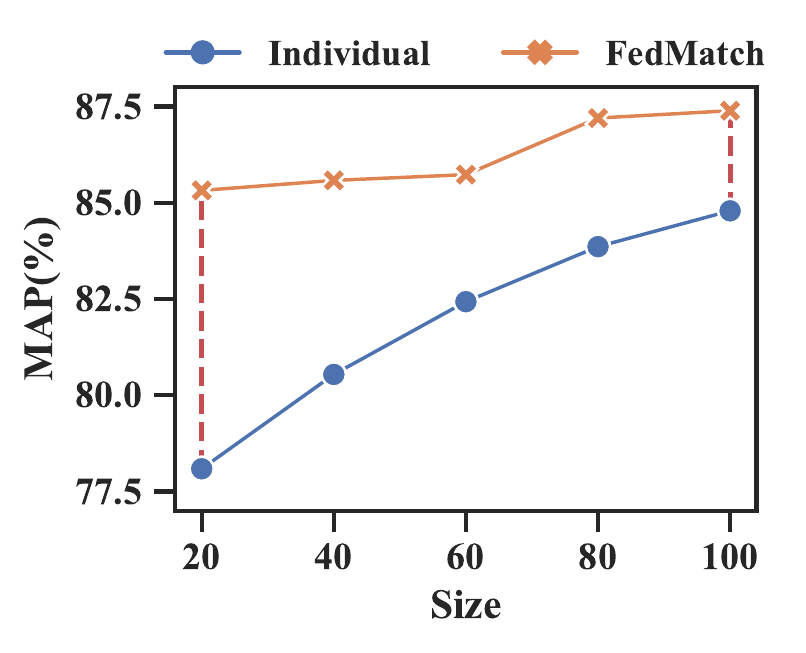}
 }
 \subfigure[MRR]{
 \includegraphics[scale=0.50]{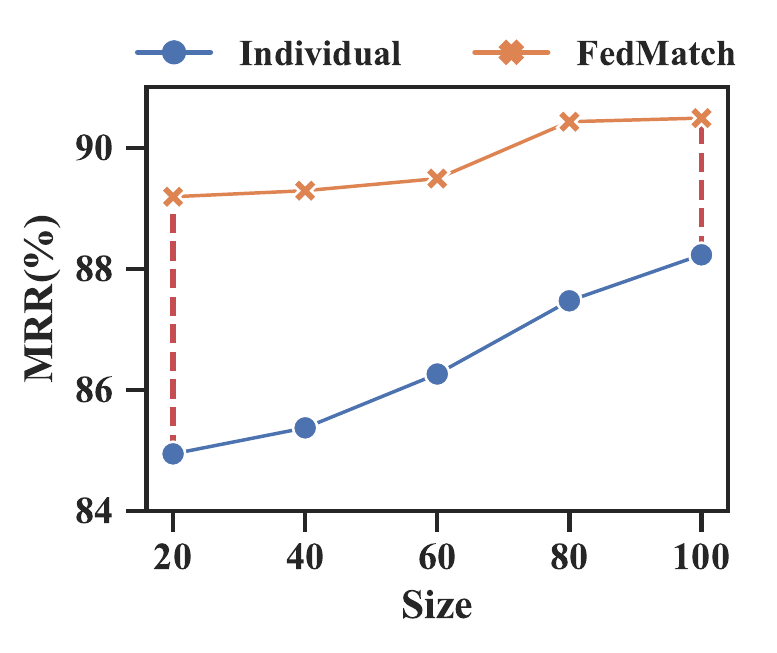}
 }
 \caption{Performance comparison of the FedMatch$_{H+LR}$ method over different sizes of training data.}
 \label{fig:size}
\end{figure}

\subsection{Impact of Training Data Size}

Here, we further explore whether the proposed FedMatch can effectively handle the data scarcity problem in each participant by leveraging the useful data of different participants.
Due to space limit, we only show the MAP and MRR results on the InQA dataset. 
We randomly select different ratios of data for model training, i.e., 20\%, 40\%, 60\%, 80\% and 100\%,   
As shown in Figure~\ref{fig:size}, we can observe that:  
(1) Compared with individual models trained on local data, the FedMatch could always achieve better performance with different ratios of data, due to leveraging the useful information from multiple participants. 
(2) The performance improvement of $FedMatch_{H+LR}$ over $BERT$ becomes more significant, as the size of labeled data on each participant decreases, i.e., the data scarcity problem in single participants in more serious. 
For example, the MAP margin between $FedMatch_{H+LR}$ and $BERT$ is 2.60\% when the ratio of data is 100\%, while the MAP margin between $FedMatch_{H+LR}$ and $BERT$ is 7.23\% when the ratio of data is 20\%.

\section{Conclusion}
In this work, we proposed to adopt federated learning for QA, which could leverage all the available QA data to boost the model  training and remove the need to directly exchange the privacy-sensitive QA data among different participants. 
With the special concern on the statistical heterogeneity of the QA data, we introduced a novel Federated Matching framework for QA, named \textit{FedMatch}, with a backbone-patch architecture. 
By decomposing the QA model in each participant into a shared module  and a private module, it is able to leverage the common knowledge in different participants and capture the information of the local data in each participant. 
Furthermore, we built a new benchmark dataset FedQA to simulate the heterogeneous situation in the real-world scenario. 
Empirical results showed that our method can effectively improve the performance by exploiting the useful information of multiple participants in a privacy-preserving way.

In the future work, we would like to enhance the data security guarantees by adopting local differential privacy techniques and reduce the communication cost via some distilling mechanisms. 
Besides, it is valuable to apply FedMatch to other tasks with the problem of data heterogeneity, such as personalized search. 

\begin{acks}
This work was funded by the National Natural Science Foundation of China (NSFC) under Grants No. 62006218, 61902381, 61773362, and 61872338, Beijing Academy of Artificial Intelligence (BAAI) under Grants No. BAAI2019ZD0306, the Youth Innovation Promotion Association CAS under Grants No. 20144310, 2016102, and 2021100, the Lenovo-CAS Joint Lab Youth Scientist Project, and the Foundation and Frontier Research Key Program of Chongqing Science and Technology Commission (No. cstc2017jcyjBX0059).
\end{acks}


\bibliographystyle{ACM-Reference-Format}
\bibliography{sample-base}


\end{document}